# Digital Twin: Where do humans fit in?

Ashwin Agrawal, M.S.[1]; Robert Thiel, M.S.[2]; Pooja Jain, M.S.[3];
Vishal Singh, Ph.D.[4]; Martin Fischer, Ph.D.[5]

[1]Department of Civil and Environmental Engineering, Stanford University, Stanford, CA, USA (Corresponding author), Email: ashwin15@stanford.edu
[2]Vice President, WSP USA, Morristown, New Jersey, USA, Email: bob.thiel@wsp.com
[3]Vice President, WSP USA, San Francisco, California, USA, Email: pooja.jain1@wsp.com
[4]Associate Professor, Centre of Product Design and Manufacturing, Indian Institute of Science, Bangalore, India, Email: singhv@iisc.ac.in
[5]Professor, Civil and Environmental Engineering, Stanford University, Stanford, CA, USA, Email: fischer@stanford.edu

## ABSTRACT

Digital Twin (DT) technology is far from being comprehensive and mature, resulting in their piecemeal implementation in practice where some functions are automated by DTs, and others are still performed by humans. This piecemeal implementation of DTs often leaves practitioners wondering what roles (or functions) to allocate to DTs in a work system, and how might it impact humans. A lack of knowledge about the roles that humans and DTs play in a work system can result in significant costs, misallocation of resources, unrealistic expectations from DTs, and strategic misalignments. To alleviate this challenge, this paper answers the research question: *When humans work with DTs, what types of roles can a DT play, and to what extent can those roles be automated?* Specifically, we propose a two-dimensional conceptual framework, Levels of Digital Twin (LoDT). The framework is an integration of the types of roles a DT can play, broadly categorized under (1) Observer, (2) Analyst, (3) Decision Maker, and (4) Action Executor, and the extent of automation for each of these roles, divided into five different levels ranging from completely manual to fully automated. A particular DT can play any number of roles at varying levels. The framework can help practitioners systematically plan DT deployments, clearly communicate goals and deliverables, and lay out a strategic vision. A case study illustrates the usefulness of the framework.

# 1. INTRODUCTION

There is little doubt that the capabilities of Digital Twins (DTs) across different use cases continue to improve, especially with recent advances in Artificial Intelligence (AI) and Machine Learning (ML). DTs are becoming more intelligent [1] and are capable of doing things only humans could do previously, such as representing and interpreting data (e.g., DT showing the health status of factory equipment) as well as making decisions and executing actions (e.g., DT generating "what-if" scenarios based on equipment's health and contacting the maintenance team if required).

With increased capabilities, DTs are not merely an expert-centric tool but proactively make decisions, complete tasks, and adapt to changing conditions [2], dramatically changing the roles of human operators. The operators will be expected to collaborate with DTs, monitor the performance of DTs while they perform certain roles (or functions) autonomously, and complete any roles (or functions) that DTs are unable to perform. This results in a piecemeal (semi-autonomous) system in which some roles, which were traditionally performed manually by humans, are now automated by DTs, while others are still handled manually by humans.

Due to this new way of working, new types of interactions will occur between humans and DTs [3], resulting in new coordination demands [4] and increasing the importance of an intelligent and efficient role allocation between DTs and their human operators [5]. As a part of the successful deployment of DTs in practice, it becomes vital that role allocation ensures that both DTs and humans are assigned tasks that they can do well. Performing incorrect role allocation can negatively affect human performance, as observed previously in aerospace and manufacturing [6,7,8]. It can result in unfavorable cost-benefit trade-offs, reduced human monitoring, over-reliance on DTs [9], and loss of situational awareness [10]. Therefore, a rigorous understanding of this joint human-DT system becomes critical [2]. In particular, DT's roles (or functions) in a work system have to be delineated from those of humans [11,12,13,14,15,16].

However, research concerning human-DT interaction is lacking in the current literature [17,18]. [19] identifies this as one of the major research gaps in manufacturing and recommends significant research efforts to fill this gap. [2] states the need to investigate the boundaries between DT autonomy and human agency to ensure DTs have the most autonomy possible without sacrificing human agency. [20,21] review the DT literature and find a lack of research on human integration in DT deployment. [15,22] also note similar findings. [23] calls for a holistic understanding of how the roles and responsibilities of human workers have changed in designing work and work systems in Industry 4.0. Therefore, to address the gap in knowledge, this paper studies DT-human interaction and answers the research question: *"When humans work with DTs, what types of roles can a DT play, and to what extent can those roles be automated?"*

Understanding how DTs can be integrated with humans is a prerequisite to harnessing the full potential of DTs and human operators [17]. A lack of knowledge about the roles that humans and DTs play in a work system can result in significant costs [24], misallocation of resources, unrealistic expectations from the technology [9,25], and strategic misalignments [26]. [4] also emphasizes the importance of knowing what role humans and DTs play in a work system among system designers and thus states: "Examination of human performance issues is especially important because modern technical capabilities now force system designers to consider some hard choices regarding what to automate and to what extent, given that there is little that cannot be automated."

This paper thus proposes a two-dimensional conceptual framework, Levels of Digital Twin (LoDT). The framework is an integration of the types of roles a DT can play, broadly categorized under (1) Observer, (2) Analyst, (3) Decision Maker, and (4) Action Executor, and the extent of automation for each of these roles, divided into five different levels ranging from completely manual to fully automated. A particular DT can play any number of roles at varying levels.

A real-life example is presented in Figure 1 to better demonstrate the research question and the need for two dimensions in the framework. The example is about Ryan, a manager on a construction project responsible for installing and maintaining wind turbines. To prevent crane tip overs caused by soil failure or unsuitable conditions, Ryan plans to create a DT of the turbine installation site that contains real-time data on soil characteristics, soil capacity, and topography. Figure 1 shows the roles required to complete the different facets of work on the project (the first dimension of our framework). Scenario 1 outlines the baseline case where no DT is used, and humans complete all the work. DT automates (partially) certain types of roles in scenario 2, thus changing the work humans used to perform within each role. Scenario 3 differs from scenario 2 in terms of the extent of automation by DTs for different types of roles which is captured by the second dimension of our framework. More details on this case study are provided in Section 5.

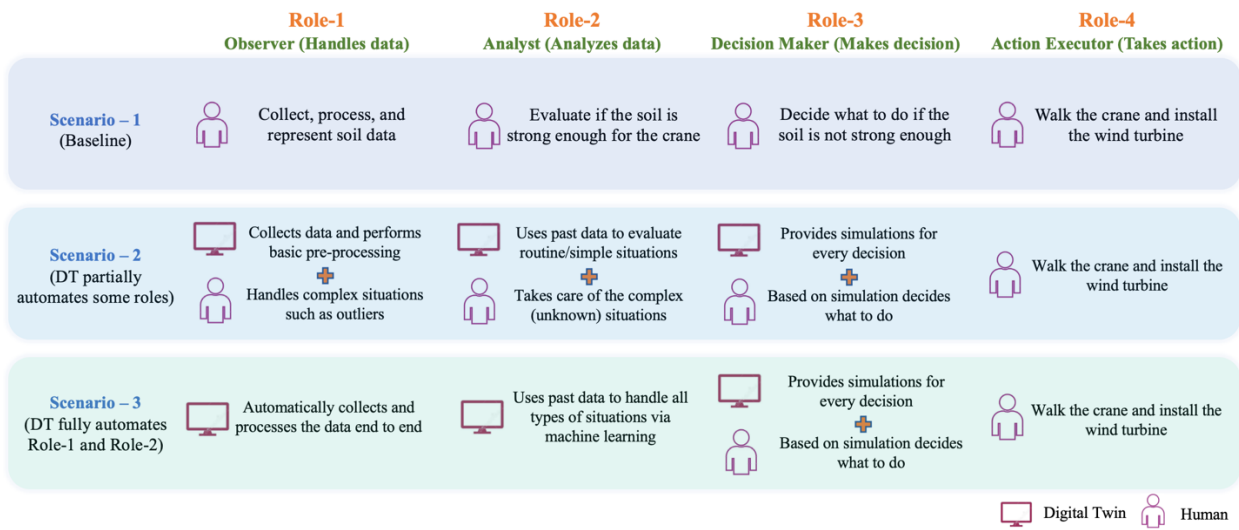

*Figure 1: A real-life example of how DT can perform various roles and the extent to which automation can occur within each role and its impact on humans*

Section 2 reviews the background of DTs, and Section 3 details the research and validation method. The proposed LoDT framework is introduced in Section 4. This is followed by showcasing the relevance of the LoDT framework in a real-life case study in Section 5. Section 6 concludes the paper by stating the contributions and the implications for industry.

## 2. BACKGROUND

This section first reviews the research context for DT in sub-section 2.1 and presents definitions, applications, and perspectives existing around the DT concept in industries such as manufacturing, aerospace, and building construction. It then focuses on the studies most relevant to this work's focus area, DT-human interaction and LoDT, in subsections 2.2 and 2.3. Finally, the gap in knowledge is summarized in sub-section 2.4.

### 2.1 DT in literature

This section reviews the history of DTs, different definitions of DTs, application areas for DTs, and the process of creating DTs.

#### 2.1.1 History of DT

NASA's Apollo program used a physical twin, a predecessor to DT, to simulate the precise conditions of an inflight test and train astronauts [27]. To construct a physical twin of every asset in the real world would be extremely costly and almost impossible. So, the idea of physical twinning was extended to create a twin digitally. The concept of digitally twinning an asset that mirrors actual operating conditions but is also practical and less costly is precisely the motivation for the DT concept [12].

The term DT was first introduced as a "Mirrored space model" in 2003 as a part of a university course on Product Lifecycle Management [28]. It was originally described as a digital information construct of a physical system linked with the physical system in question. DT is considered to be consisting of three parts [29], as shown in Figure 2: (1) the physical entity, (2) the digital entity, and (3) the data flow between them. Any change in the state of the physical object leads to a change in the state of the digital object and vice-versa. The term DT was not yet coined and was only associated with the above concept of the "Mirrored space model" in 2011/12 by Vickers and Grieves [28]. After this, the first formal definition of DT was coined by NASA in 2012 as [30]: "an integrated multi-physics, multi-scale, probabilistic simulation of a vehicle or system that uses the best available physical models, sensor updates, fleet history, etc., to mirror the life of its flying twin."

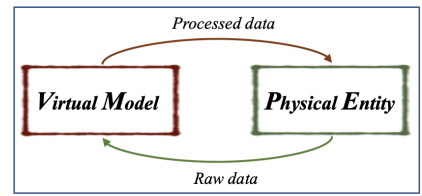

*Figure-2: The Digital Twin Paradigm*

#### 2.1.2 DT description in literature

Despite the DT concept being in the industry for a long time, it remains ambiguous [31]. It could refer to a highly sophisticated predictive and prescriptive model for some [30,32]; for others, it can be a simple digital representation [33,34]. [35] states that "literature has referred to DT as a virtual or digital model, layout, counterpart, doppelganger, clone, footprint, software analog, representation, information construct, or simulation of its physical counterpart."

In spite of these inconsistencies in DT definitions in the literature, there are a few core characteristics of DT that appear to be shared by all definitions, namely, DT should have the intelligence to sense and understand the environment (enabled by the data flow from a physical entity to a virtual model), and the agency to change the environment (enabled by the data flow from virtual model to physical entity) [36]. Similar observations have also been provided by [2,37], where the authors emphasize the importance of autonomy (delegating control) and bi-directional data flow in a DT.

Therefore, for the scope of this paper, we adopt a working definition of DT proposed by [38]: *"A digital twin is a digital replica that imparts some form of intelligence and agency into the entities being twinned to achieve the desired function."* This definition has been adopted for the following reasons:

1. It captures the core characteristics of DTs: intelligence and agency. Furthermore, intelligence and agency become intuitive and natural in analyzing the types of roles that DTs can play in a work system, which is the focus of the paper since, with changing roles, intelligence and agency must also change.
2. The definition is broad and agnostic to industry sectors and applications, conforming to the suggestion by [39,40], who notes that a good definition of DT must not include objectives (or applications) of DTs as they are always emerging and potentially infinite.

Our paper provides this working definition of DT for the purpose of answering the research question. We also acknowledge that the plethora of definitions of DT in the literature results in confusion among practitioners and researchers regarding different terminologies related to digitalization [35]. The LoDT framework, to some extent, helps alleviate this confusion. More discussion on this is provided in Section 6, where we argue that none of the proposed definitions of DT is incorrect. In their definitions, different researchers have partially proposed different roles that they envision a DT to play, and all of them unify when we consider the complete set of roles DTs can perform within the LoDT framework.

### 2.1.3 Application areas of DT

Owing to its usefulness, DT has spread to many industries and has emerged as a key concept for digital transformation in the construction [41,42,43,44], manufacturing [20,45,46], and aerospace [47,48] industries. It has shown a significant promise to increase productivity, reduce operational costs, improve safety, and optimize asset sustainability in these industries [49]. In the manufacturing industry, DT has been used for a variety of applications, including product lifecycle management [50], production planning and control [51], and process redesign [27]. In the construction industry, DT has been used in building operations and maintenance [52], fragility assessment [53], construction safety [54], performance-based smart contracts [55], and building energy simulations [56]. A recent survey by McKinsey estimates 70% of manufacturers to use DTs by 2022 [57].

## 2.1.4 Design and creation of DT

Current literature in manufacturing systems suggests that DT can be built in two ways [58]: (1) System-based and (2) Data-based.

In system-based DTs, the goal is to represent the actual manufacturing system as close as possible so that it can act as a single source of truth containing all existing links between the components at logical, physical, and functional levels [59]. Therefore, to create a system-based DT, knowledge of all technical details and data sheets for the individual components of the manufacturing system is needed. Such a DT is especially helpful for simulations to understand why an object behaves in a certain way under certain circumstances, as well as understand a system's behavior under different environmental conditions to which it is not yet exposed. In summary, a system-based DT is difficult and resource-intensive to construct but provides very comprehensive insights about the entity being twinned. [27,60] use system-based DTs for designing products and services in manufacturing and production engineering. [61] reviews the literature and finds that system-based DT has been used to design function-structure-behavior-control-intelligence-performance of manufacturing systems.

On the other hand, data-based DTs focus on creating a DT by collecting the data required to fulfill a specific function using sensors and therefore offers less insight into the interior of the manufacturing system. However, the advantage is that it is easier to create, and not all technical information of the system is required to create it. Due to the ease of use, data-based DTs are gaining more popularity and are being used to design many manufacturing systems, such as forecast models and models for predictive maintenance [62]. The first step to creating a data-based DT is to ensure that all required data is accessible using the Internet of Things (IoT). In the second step, models, analyses, and functions are created using big data analytics and machine learning. The last step is the creation of user-specific applications, which provide important information to the user, contributing value to them [58].

DTs can also be created with the combination of both approaches. The data model of the data-based approach is combined with the system model of the system-based approach, thus retaining the advantages of both approaches.

## 2.2 DTs working with humans

When humans work with DTs, two steps are involved: Allocating work or tasks so that roles are defined, and there is a clear understanding of who will do what; and once the role definition is clear, ensuring a smooth human-DT interface so that both parties can successfully collaborate and complete the task they have been assigned [63].

The idea of role allocation, which is the focus of this work, is related to concepts like (1) work allocation between humans and computers (or machines), (2) work allocation between humans and robots, and (3) human integration with Cyber Physical Systems (CPS), human-in-the-loop CPS, and human-in-the-loop Industry 4.0. Therefore, the concepts discussed in these related fields are relevant to our research. Hence, we provide a summary of the existing literature in these fields as a precursor to our framework.

Fitts published one of the first works describing work/role allocation in 1951, listing "what men are better at" and "what machines are better at," each entity being assigned the role that they are best at [64]. The list is, of course, not up to date because modern technology has outpaced humans in several categories. Several attempts have been made since then to create taxonomies that identify scenarios where humans and machines (or computers/robots) can collaborate in different ways. A popular example of such a taxonomy is by [3] in 1978, who proposed a *10-level automation* list, where humans do all the work in level-1 and computers do all the work in level-10. On similar lines, several authors [37, 38] have proposed *levels of automation* frameworks to allocate work between humans and computers. Though these studies have been criticized for not providing proper guidance on its use [67], they provide important insights into the roles that a DT can play when working with humans, as discussed in Section 4.1.

[68,69,70] propose taxonomies of the human roles in human-in-the-loop Cyber Physical Systems (CPS). These taxonomies identify four roles: Data acquisition, State inference, System influencing, and Actuation. These taxonomies have three major shortcomings: (1) although the authors identify the roles, they do not indicate how much of the role is performed by humans and DTs (extent of automation) because, for example, the same role of data acquisition can be performed partially by both, (2) these taxonomies use a lot of technical jargon which makes them difficult for practitioners to understand, use, and communicate. [71] emphasizes the importance of jargon-free frameworks for more practical use, and (3) these taxonomies are less useful for practitioners as they do not indicate which roles or extent of automation might be easier to achieve in practice. This is particularly important in the AEC industry because the companies are more tentative about adapting new technologies. Therefore, role allocation and human-in-the-loop in DT/CPS/Industry 4.0 continue to be active areas of research [72].

The next step after assigning roles to humans and DTs is to ensure that the human-DT interface is smooth. Therefore, it is necessary to define logical activities that support interaction with humans and user goals [73]. Of course, the type and amount of human-DT interaction (and, therefore, the activities supporting the interaction) will vary based on the task allocation between them. For example, a situation where humans are actively involved with DTs in completing a task would have much higher interaction than a situation in which they are just supervising DTs.

This paper focuses on the work/role allocation between humans and DTs. Therefore, the human-DT interface becomes an area for future research that can build on this work. We provide a brief overview of the existing literature on human-DT interfaces, but a full review is beyond the scope of the work.

The idea of the human-DT interface is closely related to Human-Computer Interaction (HCI) and Human-Machine Interaction (HMI), which are active areas of research in computer science and manufacturing. It focuses on the interfaces between humans and DTs and tries to ensure smooth communication, cooperation, and interaction between them [74]. There are three main areas of focus [69]: (1) engaging the human through a natural, understandable, and unobtrusive interface,

(2) the attentional resources required to perform a collaborative task, and (3) human factors such as fatigue, stress, demotivation, trust in automation, ergonomics, and optimal workload levels [75]. Some of the main technologies supporting the human-DT interface include virtual reality, augmented reality, haptic interactions, gesture recognition, and voice recognition [76]. General Electric has developed a graphical interface to display, control, and manage multiple digital models [77]. Siemens has developed a human-programming interface that enables DT to interact with humans and interpret their behavior [21]. Though these pilot applications of DT interactions are motivating, very few studies exist on the interaction and collaboration of DTs [21], making this a potential area for future research.

## 2.3 Levels of Digital Twin (LoDT)

To the best of our knowledge, no existing work focuses on holistically describing the different roles that DTs might play and the extent to which those roles can be automated when DTs work with humans. However, as role allocation is linked to the categorization of different types of technological capabilities found in DTs, we summarize existing approaches and frameworks in this area.

Autodesk [78] defines the five levels of DT: (1) Descriptive twin, a visual replica of the asset, (2) Informative twin, captures and aggregates defined data, (3) Predictive twin, uses operational data to gain future insights, (4) Comprehensive twin, generates 'what-if' scenarios, and (5) Autonomous twin, acts on behalf of the users. On a similar line, [79,80] also provides levels of digital twin maturity. Although these categorizations provide the types of roles that a DT can play, they fail to capture the extent of automation for these roles. For example, a DT making predictions in an uncertain environment (e.g., Will the project finish on time?) is more context-aware and autonomous than the one making predictions for routine tasks (e.g., Will the sun rise tomorrow?), although both perform the same role (or function) of prediction.

 [81] provides a categorization based on levels of automation in a DT: Pre-Digital Twin, Digital Twin, Adaptive Digital Twin, and Intelligent Digital Twin, but does not capture the different types of roles. [37] describes three types of DT based on the level of data integration: (1) Digital Model, which is a digital representation of the physical object without any data exchange, (2) Digital Shadow, which includes one-way data flows from the physical object to the digital object, and (3) Digital Twin, which has bidirectional data flows. This categorization is based only on the data integration capabilities of a DT and thus does not identify the types of roles or the extent of automation.

## 2.4 Gap in knowledge

To harness the full potential of DTs and human operators, it is important to understand the types of roles a DT can play and to what extent the roles of the DT can be automated? A review of the existing literature indicates that there are no studies that provide this complete information. Moreover, most of the existing classifications which partially answer this question have been provided by different commercial firms, thus lacking the necessary validation and raising questions about rigor and comprehensiveness.

To alleviate this gap in knowledge, this work proposes the LoDT framework, which is an integration of the types of roles a DT can play and the extent of automation for each of these roles. The framework has been developed following the Design Science Research (DSR) methodology, and its usefulness is illustrated through a real-life case study.

## 3. RESEARCH METHODOLOGY

The LoDT framework has been developed and validated over the course of 36 months by using the Design Science Research (DSR) methodology [82]. The DSR methodology is well suited to this context as it provides the following advantages:

1. *Enables solving "wicked" design problems*: In many real-world scenarios, as in our case, the user (or client) is unsure of what they want. As a result, it is difficult to define the explicit requirements and constraints of the problem at the beginning of the project, making it difficult to find a satisfactory solution [83]. To alleviate this challenge, DSR supports developing and refining both the formulation of a problem and ideas for a solution together, termed co-evolution [84]. For our work, this is relevant because the client's requirements were not well articulated at the start of the project, and those requirements evolved as we developed the solutions, further outlined in the research objectives section.

2. *Helps design practical solutions:* DSR is applied in nature and thus not only describes and explains the problem but also helps design a solution (e.g., models, methods, or frameworks) to solve it [85]. It balances both theoretical (existing literature, frameworks, and models) and practical influences (business needs and practical deployment) [86] to make the solution conceptually rigorous and relevant in practice. This is not possible by other research methods such as surveys and questionnaires [87]. It uses an iterative, constructive, and pragmatic approach [83] to construct and evaluate the solution, enabling researchers to receive constant feedback from the end-users of the framework. Again, this makes DSR suitable for our research because we also intend to create a practical framework (see non-functional requirements of the framework described later in the paper).

The research methodology is illustrated in Figure 3. As per the DSR methodology suggested by [88], the problem was identified and validated, and specific objectives for the research were defined. This was followed by the design and development of the framework. Iterative demonstrations and evaluations of the artifact were conducted through extensive presentations to 11 experts (over 50 hours) and two public presentations, including over 40 experts in each presentation. The following paragraphs detail each of these steps.

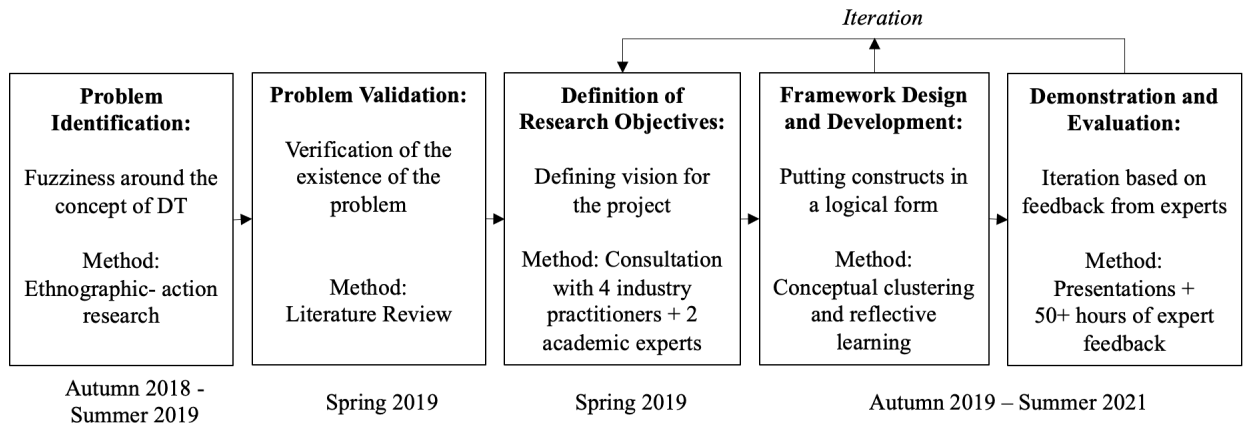

*Figure 3: Research Methodology*

**Problem identification, problem validation, and definition of research objectives:** Based on an ethnographic-action study conducted by [11], the problem of practitioners struggling to understand how DT fits into a work system was identified. A literature review of the application of DT in various industries further reinforced the gap in knowledge about human-DT integration, as summarized in Section 2.

To set the specific objectives and requirements for this research [89], the authors met with four industry practitioners and two academic experts. In the first iteration, the requirements for the framework were defined as: "Logically explain the complexities of available DTs and the functions they support." After initial prototyping, inspired by "five-levels of vehicle autonomy" [90], this transformed into: "Develop DT hierarchy for construction, similar to levels of autonomous driving." After a few more rounds of framework prototyping, it became clear that DTs would not be working alone but with humans. Therefore, it became vital to consider the human perspective as well. Hence, the final functional and non-functional requirements (following [91]) for the LoDT framework were defined as follows:

*Functional requirements:* (1) Provide a classification of the various types and levels of DT available in the industry and their impact on humans, (2) provide practitioners with an outline of the level of DT their company currently has and the level that they hope to achieve in the future, and (3) explain what each level offers, as well as what it takes to make it in terms of difficulty and resources.

*Non-Functional requirements:* (1) Intuitive to understand, (2) quick and easy to use, (3) jargon-free for quick recall and easy communication, (4) insightful in practice, and (5) "general enough" for different industries to adapt.

**Design and development:** The research process described by [86] was followed for the framework development. Following the functional requirements, the intent of the initial design was to (1) describe and explain different types, ideas, and technologies that are commonly termed DT and (2) arrange them in the form of a hierarchy that relates to the way humans work.

Many other forms of taxonomies and hierarchies were studied in the contexts of level of automation [4,66], levels of autonomous driving [90], and types of intelligence [92] to inform the initial design.

New constructs for representing the hierarchy were identified and described through a process of inductive inference, conceptual clustering, and reflective learning [93,94,95]. Specifically, the intention was to answer questions similar to: What are the fundamental elements and themes that characterize a DT? What are the specific functions, roles, and applications provided by DT? What type of intelligence capabilities should be present in a DT? These constructs were then assembled into a logical artifact (the framework) that could visually present a clear and concise picture of LoDT.

The artifact was then demonstrated to experts (see Table 1), and multiple iterations of design and development were carried out. The demonstrations to the experts were semi-structured in nature, starting with the presentation of the latest iteration of the framework and noting down the specific feedback before turning to a broader discussion. This step was performed to ensure that researchers were able to gain first-hand reflection of the experts in an unbiased way. Specifically, keeping in mind the non-functional requirements, the experts were asked to comment on: (1) the elements of the framework that made sense to them (to understand the strong points of the framework), (2) the elements that need further research (to understand the weaker points of the framework), (3) whether they think the framework would help them in practice (to ensure that practitioners found value in the framework), and (4) whether they would be comfortable using the current version of the framework in practice (to ensure that the framework is simple enough to be used in practice).

In the broader discussion, the experts were asked to describe a few of the success and failure anecdotes that they had experienced while deploying DT in their firms. The researchers then took a retroductive approach [96,97] and postulated if the framework captured the necessary details and provided insights to be helpful in the described situation by the expert, thus checking if the framework met the non-functional requirements. This step was performed in order to (1) ensure the framework is relevant in practice and (2) identify the gaps in the framework by working through specific scenarios. The latter helped improve the framework in future iterations. Some of the comments by the experts which were pivotal during the framework development are summarized in Table 2.

**Demonstrations and evaluations:** The framework was evaluated by carrying out demonstrations to various practitioners and academic experts over the course of three years and in two ways: (1) major milestone presentations and (2) regular feedback sessions. The feedback from each demonstration was incorporated into the next iteration of the framework.

A major milestone presentation was held on a yearly basis (for the years 2019 and 2020) and consisted of 40+ industry and academic experts as the audience. The state of the framework used in these presentations has been documented digitally and can be provided on request for understanding the trail of the research process. The major milestones acted as a good sanity check for the progress of the framework and ensured that the broad-level story was well received by the experts.

For more granular feedback, regular feedback sessions were conducted between 2019-2021 with 11 experts (shown in Table 1) with expertise in technology, strategy, and the working of the industry. A total of 40+ meetings, clocking over 50 hours, were conducted to ensure the relevance and rigor of the framework. Multiple meetings were conducted with the experts to ensure they had sufficient time to understand, reflect, and comment on the framework and that their feedback was properly accommodated. The iterations were carried out until a 'theoretical saturation' [98] was reached, i.e., no new or relevant feedback emerged from the demonstrations.

| | Expert Code | Background/Role | Experience (Years) | Number of Meetings | Total hours of interaction |
|---|---|---|---|---|---|
| Technology experts | A | Expert and researcher in use of AI | 30 | 10 | 10 |
| | B | Expert in deploying new technologies | 25 | 1 | 1 |
| | C | Researcher and founder of a technology startup | 22 | 2 | 1 |
| Strategy experts | D | Expert and researcher in innovation management | 11 | 3 | 3 |
| | E | Researcher in management and technology | 4 | 5 | 7 |
| | F | Expert and researcher in technology strategy | 40 | 2 | 2 |
| Industry experts | G | Part of Senior Management in an engineering firm | 24 | 8 | 8 |
| | H | Project Manager on a $1B+ construction project | 28 | 4 | 10 |
| | I | CEO of consulting and innovation firm in building construction industry | 24 | 3 | 3 |
| | J | Part of Senior Management in a multi-national construction firm | 18 | 4 | 3 |
| | K | Management executive in an engineering consultancy firm | 34 | 1 | 2 |
| *Total of 50-hours of interaction spreading over 36 months with 11 experts (average experience of over 23 years)* | | | | | |

*Table 1: Demonstration and Validation of the framework, profile of the participants*

| Sample quotes from the Experts | Consequential insights gained for the framework |
|---|---|
| "In the Digital Twin: "What is Digital?", and "What are we twinning?"" "Form of the DT should follow its function." | Forced the researchers to think about the functions that DT can aid with, shifting away from taking a technology-centric view. Ultimately led to the formulation of four types of roles a DT can play (first dimension of the framework). |
| "The progression of cognitive abilities isn't linear. Like the decision to predict if the sun would rise tomorrow is easier compared to predicting who will win the football match, even if both are predictions." | Led to the realization that a second dimension representing the extent of automation is needed. |
| "Some tasks are easy and always follow a repetitive procedure. On the other hand, some tasks are very dynamic and change every time. These are inherently difficult to automate." | Played an important role to introduce the concept of Learning (routine v/s non-routine tasks) (see section 4.2 for details). |
| "We shouldn't always aim for a DT to complete 100% of the task. Even if the DT completes 80%, and the human completes the rest 20%, it would still be very helpful." | Led to the introduction the concept of autonomy (see section 4.2 for details). |

*Table 2: An excerpt of quotes from experts that led to consequential changes in the framework*

# 4. LEVELS OF DIGITAL TWIN FRAMEWORK: AN INTRODUCTION

This section introduces the LoDT framework. It is represented in two dimensions as shown in Figure 4: (1) types of roles that a DT can play (columns), and (2) extent of automation (rows). The two dimensions are respectively introduced in Sections 4.1 and Section 4.2. Section 4.3 provides suggestion for how to use the framework and lists frequently asked questions while using the framework.

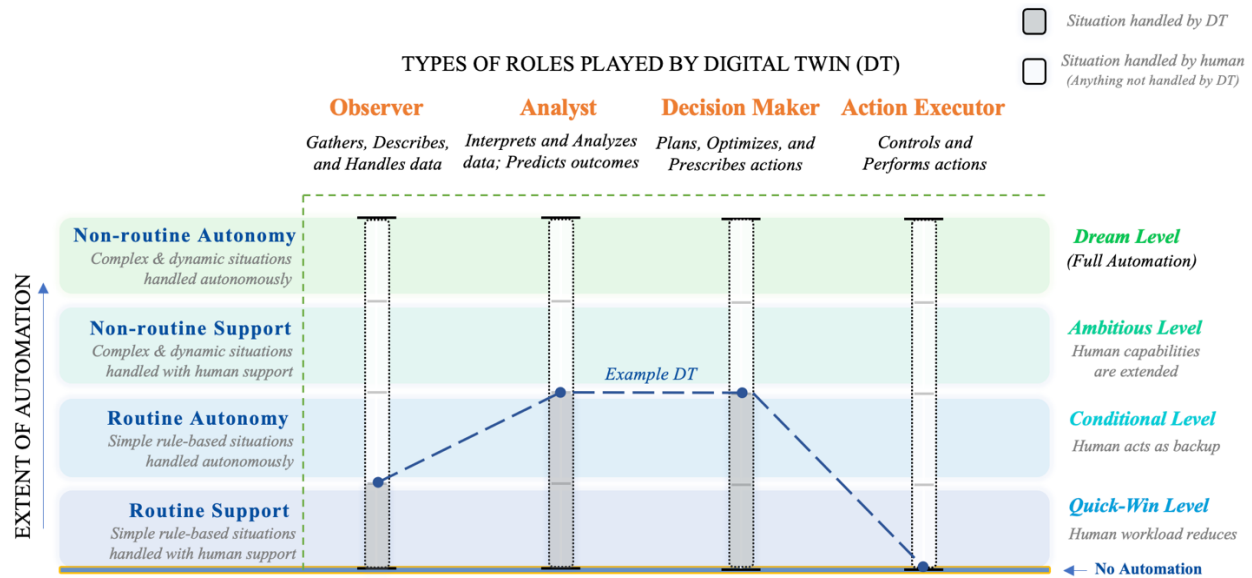

*Figure 4: Levels of Digital Twin framework. In the example shown with dotted lines, the DT would support data collection and observations in routine situations, perform routine data analysis and decision making autonomously, but not perform any actions. Humans would handle data collection, analysis, and decision making in non-routine situations. Actions in all situations would be performed by humans.*

## 4.1 Types of roles that a DT can play

A role is defined by the Oxford dictionary as "the functions assumed, or the part played by a thing in a given situation."

To identify the roles that DTs can play when working with humans, we took a two-step bottom-up approach. We began by identifying the functions that one would expect (or envision) DTs to perform. Then we identified the roles by grouping those functions into higher-level categories.

Therefore, in the first step, a list of expected functions from DTs was compiled based on a literature review and expert interviews. A total of 80+ distinct functions were identified in the first iteration. For example, DTs can fulfill functions like data representation, simulation, prediction, or automated monitoring.

Our next step was to group these functions into roles responsible for handling them. Initially, clustering revealed that some functions, such as visualization, description, and retrieval, are similar to basic cognitive functions in humans and are primarily used at the beginning of a task to understand the situation. Then there are others, such as prediction, simulation, and optimization, that are much more complex and are required toward the end of the task to make decisions and perform actions. From multiple rounds of clustering and speaking with experts, a common theme emerged that functions should be clustered according to the sequence in which they are needed to complete a task.

Let us look at our original example to illustrate this further. Given two very different tasks that Ryan might encounter, such as constructing an airport versus walking a crane, the basic sequence of roles that would be needed to complete the task is the same: Collecting, representing, and handling the data (i.e., the role of an Observer), manipulating and comprehending the information from the data (i.e., the role of an Analyst), making decisions based on that information (i.e., the role of a Decision Maker), and finally implementing the decision in the form of an action (i.e., the role of an Action Executor).

Our findings led us to adopt a four-stage view of the roles humans use to accomplish tasks. This view has been used by several authors, including [4,99]. These four categories can provide a higher level of abstraction of the roles that DTs can play: (1) Observer, (2) Analyst, (3) Decision Maker, and (4) Action Executor.

We have grouped some of the DT functions as gathered from the experts into these four categories (roles) in Table 3. Of course, these four categories are almost certainly a gross simplification of the range of roles that DTs can provide. However, our goal here is not to debate the theoretical abilities of a DT but to provide a structure that is relevant in practice. In this respect, the conceptualization shown in Table 3 is a good starting point, and it is shown to be relevant through our case-study detailed in Section 5.

| Observer | Analyst | Decision Maker | Action Executor |
|---|---|---|---|
| • Sensing | • Analyze and Monitor | • Optimization | • Physical action |
| • Representation | • Pattern recognition | • Simulation | • Actuation |
| • Description | • Prediction | • Forecasting | • Communication |
| • Visualization | • Interpretation | • Prescription | • Control |
| • Retrieval | • Perception | • Planning | |
| • Basic manipulation | • Diagnosis | | |

*Table 3: Example functions expected to be carried out by different roles*

**Observer:** The role involves obtaining, manipulating, and presenting the data of the physical world by perceiving the status, attributes, and dynamics of relevant elements in the environment. Therefore, sensing, acquiring, representing, digitizing, and communicating the data gathered from the physical world are the central functions of this role, analogous to the way humans sense the environment through their eyes, nose, ears, and other senses. An example would be Ryan using sensors to acquire data about the soil conditions and the DT helping the human operator to perform basic pre-processing.

**Analyst:** In this role, the raw data is analyzed and interpreted in light of the target goal and the preferences of the operator in order to make a judgment about the state of the physical world. Knowing what is good and what is bad is crucial.

For example, after getting all the data about the soil conditions, Ryan's DT needs to evaluate if the soil has enough bearing capacity or not. For this role, therefore, pattern recognition, monitoring, interpretation, and prediction (of missing information) are the key functions needed to be performed by DTs. Recent research in AI has focused on improving machine perception in areas such as image recognition, speech detection, and natural language processing. This allows machines to understand the information gathered from their surroundings more effectively.

**Decision Maker:** This role takes as input the analyzed information and helps make decisions. In other words, DTs need to exhibit a 'rational or utilitarian' behavior and hence plan for the actions which help it to achieve the maximum utility. Sometimes the 'rational' decisions can be straightforward, like selecting a single or a set of actions to get the desired outcome. Sometimes it would be more tricky and might require searching and planning. The decisions involve consideration of the future: "how the world evolves?", "what will happen if I do such and such?" and "what will give the maximum happiness?".

Therefore, simulation, forecasting, and optimization become critical functions for this role. Recent works in machine learning, automated decision support, and reinforcement learning have focused on improving practical decision-making capabilities for machines. As an example, suppose the DT helps Ryan to decide what to do when the soil capacity is not sufficient to walk the crane. As one might imagine, this is not a trivial question. There are several possible options, such as changing the path of the crane, using a different crane, or strengthening the soil. The DT would need to consider the least costly and quickest option.

**Action Executor:** This role can sometimes involve moving away from the cognitive domain for a DT and entails implementing the actual action based on the decision taken. The action can involve interacting and acting upon the physical environment using actuators. Advancements in robotics have greatly aided the implementation of actions by machines. For example, DT could help Ryan remotely control the crane.

It is important to note that a DT can perform one or many of these roles. Using Ryan's example, if Ryan's DT only handled data, humans would be responsible for the rest of the task (analyzing information, making decisions, and taking actions). In contrast, if Ryan's DT is only able to make decisions and perform actions, humans would be required to observe and analyze data. Essentially, any role that a DT cannot perform needs to be done by humans.

## 4.2 Extent of automation

To formulate the hierarchy for the extent of automation for each role of DTs, we start by asking: "to what extent can a given role be automated through a DT?". The answer depends both upon (1) autonomy (DT's level of independence in each role) and (2) learning ability (DT's ability to adapt to dynamic environments in each role).

Naturally, the higher the automation, the more DTs will be able to complement human workers in a specific role by working independently and adapting to changing circumstances. Thus, two dimensions are identified: (1) Autonomy and (2) Learning, through which, to a large degree, the extent of automation for each role of DTs can be expressed and compared. [67] also identifies the importance of task entropy (which is correlated to learning ability) and autonomy for identifying long-term trends in the extent of automation for various tasks. [2] mentions the three important criteria for future DTs: autonomous, context-aware, and adaptive (like learning), all of which align with our formulation.

In the following paragraphs, first, the dimensions are explained, and then the hierarchy for the extent of automation for DT roles is described.

**Autonomy:** Autonomy determines how involved humans are with DTs to complete one or several of the functions for a specific role. A fully autonomous DT completes all the functions for a specific role without any human involvement. A partially autonomous DT assists the humans and thus reduces the workload associated with manually completing the role. Several taxonomies in the human-machine automation literature [65,100] have been proposed based on the level of autonomy (control), with higher levels representing increased autonomy of computers over human action and, thus, higher levels of automation. Based on autonomy, the LoDT framework differentiates the extent of automation as follows: with human support (partial autonomy) versus completely autonomous.

**Learning:** Learning enables DTs to improve their performance on future tasks based on past performance and observations of the world, similar to the way humans learn and grow intellectually over time. Therefore, this increases the automation of each role for DTs since now the DT can perform functions that require cognitive flexibility and adaptability, as well as situations where rules are not completely understood beforehand or where no correct solution exists. Based on learning ability, the LoDT framework differentiates the extent of automation as follows: routine situations (well-defined, predictable situations that can be solved through explicitly programmed instructions) and non-routine situations (no explicit instructions are available or algorithms must adapt to some degree of dynamic change, thus requiring learning). [4,101,102,103] have also highlighted the importance of learning and differentiated between routine and non-routine tasks according to learning ability.

As one might expect, DT's highest level of automation would be for each role to work autonomously, handling non-routine situations, just like humans. While attempting to automate every role of DTs to the highest extent possible might be an envisioned benchmark, in most cases, it may not be necessary, practical, or even desirable due to a lack of technical capabilities, risk preferences of practitioners, legal requirements, or financial constraints. In these cases, intermediate levels of automation may be preferred so as to maximize the joint performance of humans and DTs. Therefore, a five-level measure of automation for each role of DTs based on autonomy and learning abilities is proposed in Figure 5.

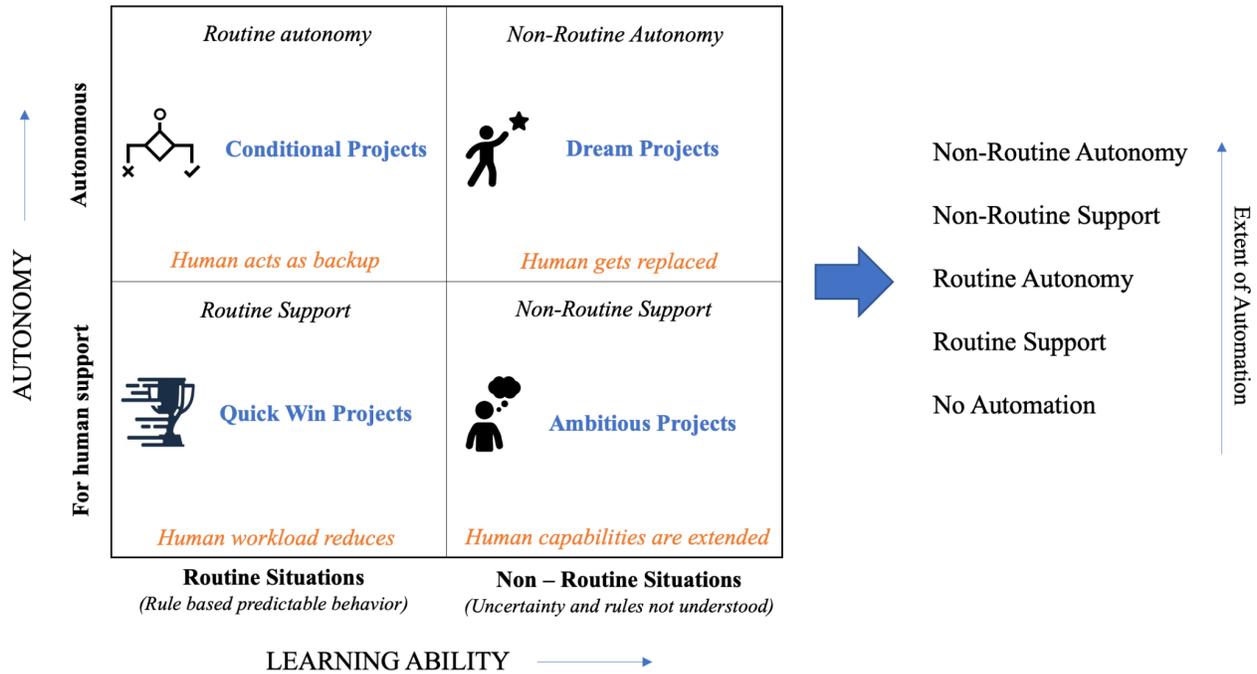

*Figure 5: Extent of automation for each role of DT*

1. *No Automation (Manual):* At this level, humans perform all the tasks without any kind of support from DT. This is the lowest form of automation that can be present in any role of DT, and it acts as default to the 2 x 2 matrix.

2. *Routine Support:* DTs assist humans in completing routine tasks by following explicitly programmed rules. The human still remains in full control and can bypass DTs based on their own judgment. From our experience, we have found that these projects are often 'quick-wins'; they are the easiest to implement with the least amount of organizational changes. By automating routine tasks, DT relieves the workload of humans [65], allowing them to focus on more important tasks that require higher creativity and intelligence.

3. *Routine Autonomy:* DTs are capable of completing routine tasks without human supervision. Autonomy in DTs creates a failure risk (such as in boundary cases), which sometimes defeats the purpose of autonomy, especially in routine situations where humans are almost perfect. Thus, from our experience, we have found that these are often 'conditional' projects. The use of routine autonomy must be justified based on the marginal benefits it provides above routine support. If the company decides to implement this level of automation, the humans act as a backup [65] to DT, performing the work in case something changes or goes wrong.

4. *Non-routine support:* By utilizing the learning element, DTs assist humans in handling non-routine tasks, i.e., situations that are dynamic or where rules are unclear. The learning process requires data, so streamlined data collection and labeling become essential. This, in turn, necessitates significant changes to the ways project organizations usually operate. There is also a need for a dynamic shift in organizational mindset towards data-driven decisions, which sometimes even questions common practice because, in non-routine situations, the correct answer may not be known in advance. Therefore, we term these projects as 'Ambitious,' requiring a significant change in the way work is performed. But on the counterpart, the benefits are also great, with human capabilities being extended [65].

5. *Non-routine autonomy (full automation):* This is the highest level of automation and that is why we term these as the 'Dream' projects for companies. DT can complete the task autonomously at the required level of performance, adapting to the task environment and improving its performance over time. We have not seen many suitable examples of a DT at this level across all roles but we are optimistic about the possibilities given the recent technological developments.

## 4.3 Frequently Asked Questions (FAQs) about the framework

This section describes the FAQ's that we got from different users of the framework during the empirical case studies conducted to validate the usefulness of the framework. These answers to the most common questions can address some of the questions that readers and new users of the framework might have, as well as provide a better understanding of how to use the framework.

1. *Is it necessary for DT to play all the defined roles, or can humans also play them?*
   It is important to recognize that the roles are needed to be performed to complete the work, regardless of whether a DT performs them or not. Therefore, any role that is not performed by a DT would be performed by humans. Figure 6 illustrates the extremes of this role distribution.

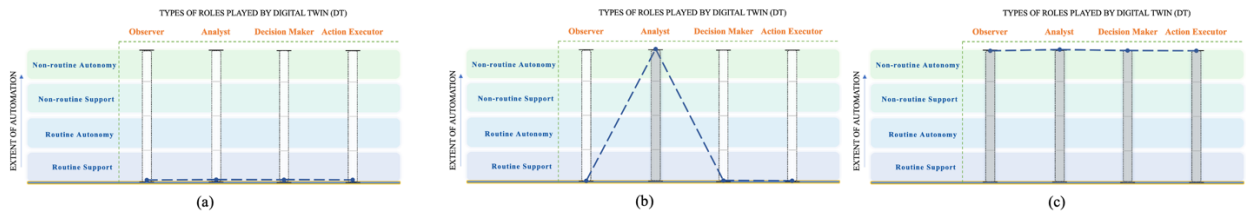

*Figure 6: Examples showing extremes for role distribution between DTs and humans. (a) DT with no automation for any roles, (b) DT with full automation for one of the roles, and (c) DT with full automation for all the roles*

2. *How can the same role be performed by DTs in some situations and by humans in other?*
   Any role in the working world involves both simple (routine) and difficult (non-routine) situations. So, if DTs are capable of handling only simple situations due to technology or any other constraints, humans would step in when a complex situation arrives. This is what the "Extent of Automation" dimension represents.

3. *How to decide what is a routine and a non-routine situation in a particular scenario?*
   We suggest thinking of routine and non-routine situations for every role. For a particular role, consider the functions that a DT would need to perform. If the function can be described by a set of explicit rules or algorithms, then it is a routine situation. It should be noted that some situations that are routine for humans are not routine for the computer and vice versa. For example, identifying a cat is routine for humans, yet it is very difficult to write explicit rules for the computer to do the same, therefore making it a non-routine task for computers. On the other hand, finding patterns in one million data points is non-routine for humans but routine for computers because it can be described explicitly through algorithms such as clustering. In the framework, we represent routine and non-routine situations from the perspective of the DT.

4. *In some cases, how can a DT have a lower level of automation for basic roles like Observer or Analyst than for advanced roles like Decision Maker or Action executor?*
   As we explained above in Question-1, it is not necessary that every role must be performed by DTs. Therefore, as long as the role is being performed and the final deliverable of that role is satisfactory, it does not matter who did it. For example, DT does not have to observe and collect the data to analyze it, as humans can collect and format it as necessary. The following example illustrates the different functions of a DT versus a human for the four roles.

   Task: A building's temperature needs to be observed and analyzed whether it is comfortable for building occupants. Then a decision needs to be made on how much to change it based on the time of day, the number of occupants, and electricity consumption. Finally, the action to change the temperature needs to be executed. Figure 7 shows how these functions can be distributed in various ways between humans and DTs.

## Example: Controlling a building's temperature using DTs

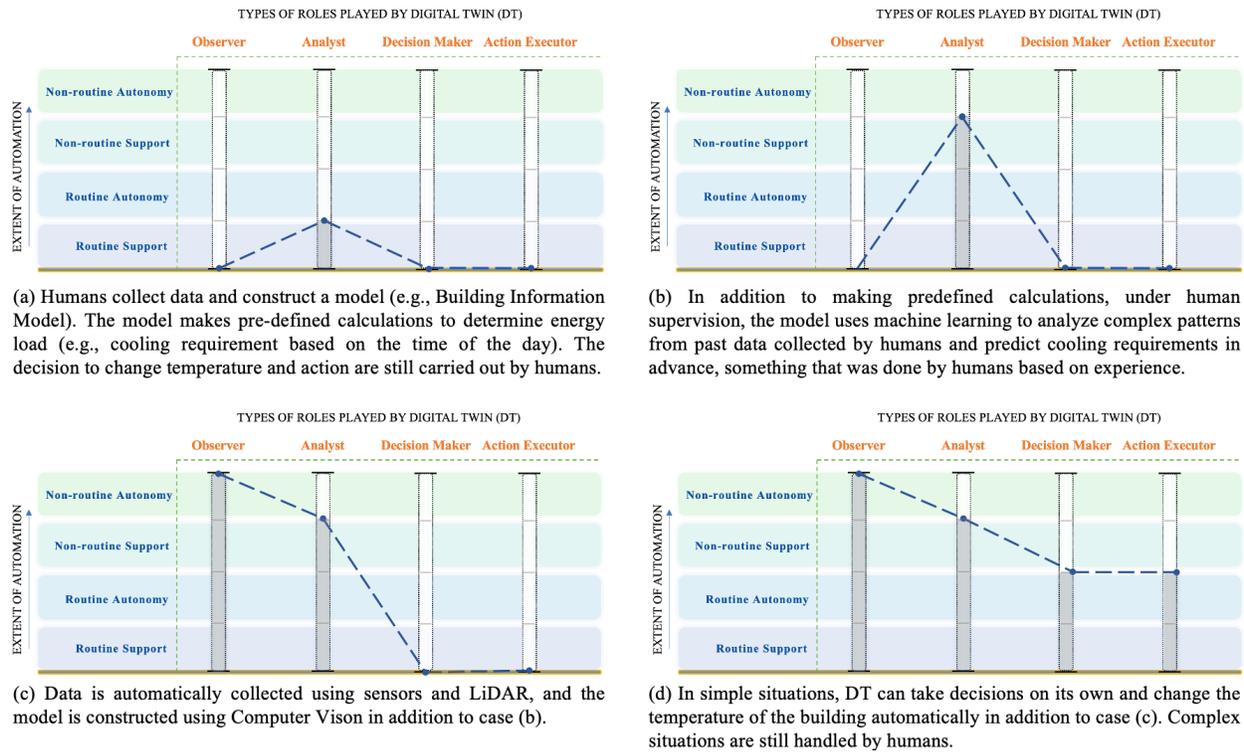

(a) Humans collect data and construct a model (e.g., Building Information Model). The model makes pre-defined calculations to determine energy load (e.g., cooling requirement based on the time of the day). The decision to change temperature and action are still carried out by humans.

(b) In addition to making predefined calculations, under human supervision, the model uses machine learning to analyze complex patterns from past data collected by humans and predict cooling requirements in advance, something that was done by humans based on experience.

(c) Data is automatically collected using sensors and LiDAR, and the model is constructed using Computer Vison in addition to case (b).

(d) In simple situations, DT can take decisions on its own and change the temperature of the building automatically in addition to case (c). Complex situations are still handled by humans.

*Figure 7: An example showing four scenarios of role distribution between DTs and humans in the context of controlling a building's temperature using DTs*

5. *Given that the framework has so many elements, how do you suggest using it?*
   While there is no "right" way to use the framework, we found that the following method provides a good starting point, especially for first-time users:

   *Step-1:* Start by defining the functions for each role required to complete the task, irrespective of who does what, as explained in Question 1. First, define the function of the Action Executor since the other roles change drastically depending on the defined action, then define the functions of Decision Maker, Analyst, and finally, Observer. For example, in the case of building a DT to maintain highways, the final action can be to fix the defects on the road, or it can also just be to detect defects on the road, depending on the scope of work. When fixing the defects, the Decision Maker must decide whether to fix it, how and when. If detecting the defects, the Decision Maker needs to decide if it is a defect and what type of defect it is. Therefore, while defining the functions for each role, users should ask themselves, to perform the defined action, what decisions do I need to make? To make the decisions, what analysis do I need? And to make the required analysis, what do I need to observe?

*Step-2:* Define what routine and non-routine situations look like for each of the defined functions under every role. For example, in the case of building a DT to maintain highways, one of the functions of the Decision Maker can be to decide whether to fix the defect. A routine situation is when DTs handle simple situations like deciding to fix the defect if it is greater than 5 inches because the defect is very large, and fixing it is the only option. Now consider the situation when the defect is 1 inch. It is not clear whether to repair the defect or leave it for later. Many factors may now need to be considered, such as how fast the defect could get worse, the available budget, and the contractual requirements. These factors might make the repair decision a non-routine situation. The DT might need to use ML to learn patterns from past data to make such decisions.

*Step-3:* Now that both the dimensions of the framework have been mapped out, the users can construct the level of DT that they currently have, as shown in Figure 5.

*Step-4:* After mapping out the current level of DT, the user can now map out the envisioned level of DT that they would like to ideally have.

*Step-5:* Finally, all the stakeholders can decide what would it take to advance the current level to the envisioned level of DT.

## 5. CASE STUDY

Using our original example, we will show how the LoDT framework helps Ryan, who wants to create a DT to prevent crane tip overs caused by soil failure or unsuitable conditions. Ryan primarily uses the LoDT framework in two ways:

1. *As a tool to plan the deployment of DT:* Ryan lacks a systematic method for exhaustively evaluating the wide range of roles of DTs and the extent to which each role can be automated. This inhibits him from selecting an appropriate DT. Therefore, Ryan uses the LoDT framework to conduct structured brainstorming and systematic role allocation between DTs and humans.

2. *To develop a strategic roadmap for DT deployment in the future:* Implementing DTs in practice is a continuous process, and therefore Ryan needs to start small and gradually progress to a DT that can perform more roles at a higher level of automation. Ryan uses the LoDT framework to understand the DT level that the firm can currently create, given its technical capabilities, as well as establish a long-term vision for where the firm might want to go.

The following paragraphs present details of how Ryan uses the LoDT framework. He starts by systematically analyzing each role and determining the extent of automation the firm will need within each role and whether they have the technical capabilities to perform it.

A. Observer: *Ability to acquire, process, and represent the data from the physical world*

While brainstorming the various extents of automation that DTs can perform for the Observer role, the following options looked more promising: (1) no automation (manual level), the data is collected by sensors and manually processed by humans, (2) routine autonomy level, DT autonomously processes the data based on simple predefined rules while humans handle complex processing. For example, DT can identify a soil type from soil composition and moisture content based on empirical rules, and (3) non-routine autonomy level, DT works in non-routine situations independently. DT, for example, learns from historical soil data and prepares soil topography maps independently, as it is very difficult to explicitly write programming instructions to create soil topography maps.

Even though the topographical data is important, Ryan feels that they can still work without it, especially since the crane operator will be handling the crane if anything goes wrong because of uneven terrain. Thus, he concludes that the routine support level of automation for the Observer role of DT will cover most of the common cases in practice and that other, less frequent cases can be handled by humans.

B.  Analyst: *Data synthesis in light of the target goal to understand what is desirable*

As the goal of this task is to walk the crane safely (without tipping), the DT in this role should be able to assess whether walking the crane is safe based on soil data and topographical maps. Some options that came up during the brainstorming session included: (1) routine support level, DTs synthesize the data and perform low-level calculations, which humans would then use to determine if the soil is capable of supporting the crane, (2) routine autonomy level, DTs calculate the soil capacity based on empirical geotechnical models that have been explicitly programmed, and (3) non-routine autonomy level, DT predicts soil capacity from historical data and can therefore handle non-routine situations such as those that occur after rain and snow or when the empirical model fails.

It is crucial to conduct a rigorous analysis of the soil capacity before walking the crane since the turbine, and its components are heavy, and a minor error in soil capacity estimation can lead to failure of the soil. Ryan notes that it takes approximately 8 weeks to manually perform the data pre-processing and calculations using finite element analysis, but the calculations are highly accurate and reliable, with a 99.999% accuracy rate. Furthermore, he stresses that even though calculation time needs to be reduced (make it almost real-time), it is extremely important to maintain these reliability standards, as any mistake could result in crane failures. Consequently, he selects the third option and sets up the target of calculating the soil capacity within 1 min of data collection, and with a reliability of at least 99.99% using ML.

C.  Decision Maker: *Evaluating choices and making a rational decision*

If the bearing capacity of the soil exceeds the crane's weight, Ryan can let the crane walk. However, if the soil is not strong enough, he must make a decision among several options, including: changing the path of the crane walk, treating the soil to make it stronger, or delaying the walk by a couple of days to let the soil settle. Some options that came up during the brainstorming session included: (1) routine support level, DT deterministically simulates all options and Ryan selects the best one, (2) routine autonomy level, DT evaluates every option and takes a 'rational' decision independently, (3) non-routine autonomy level, DT simulates options based on the uncertainty that might arise like sudden weather changes and chooses the best option independently.

Ryan feels that simulations are not necessary, at least in the short term. He, therefore, decides against it by explaining that decisions will still be made by humans.

D.  Action Executor: *Executing the action*

Finally, Ryan needs to operate the crane and install the wind turbine. The crew can complete the job manually. He can also use DTs to control the crane remotely. Finally, DTs can even control cranes autonomously at the fully autonomous level. In the short run, Ryan believes this capability is not necessary and decides not to implement any automation for action.

Compiling all the decisions, the final level of DT at the end of Iteration-1 is shown in Figure 8.

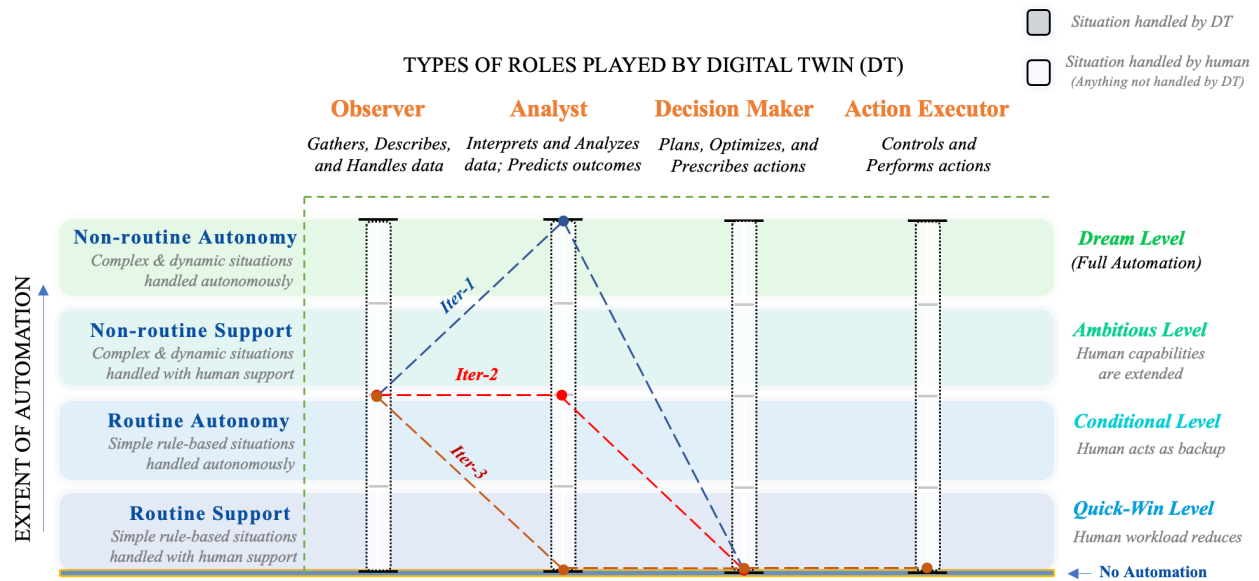

*Figure 8: Multiple iterations to select the appropriate level of digital twin*

**Iteration-2:** While working through the details, the team realizes that they lacked data for non-routine situations, such as the soil capacity data after heavy rains and snow or in uneven topographical conditions. However, the team realizes that they had a lot of data in routine situations and hence wondered if ML can be deployed in routine cases, and the human would take control in the non-routine cases. This is shown in iteration-2 in Figure 8, wherein the Analyst role, the non-routine autonomy level from iteration-1, was dropped to the routine autonomy level.

**Iteration-3:** The developers inform the team that the best performance that ML models can possibly achieve would be nowhere close to the required reliability of 99.99%. As this was a non-negotiable target due to the safety issues, the team had to drop this idea as well and finally decided to go with only using DT to observe and handle data for routine situations (See Figure 8, iteration-3). They found that the time to calculate the soil capacity decreased significantly from about 8 weeks to 2-3 days. Although this is not the desired target that Ryan had in mind, this was the best possible solution that the team would be able to achieve, given the resources in hand.

In summary, Ryan uses the LoDT framework as a planning tool to conduct structured brainstorming sessions with the goal of comprehensively evaluating the role allocation between DTs and humans rather than coming up with unstructured suggestions. Additionally, he provides a roadmap on how DT can be deployed using the LoDT framework. In the short term, because of the lack of resources, DT should only be used for routine data observation and handling, even if it does not lead to the real-time calculations, they envisioned earlier. He continues that in the future, the company could develop a DT that analyzes non-routine situations by using ML models. To do that, the company should start collecting soil capacity data in uncertain conditions, like after rain and snow, and also start recruiting people who can drive the data-driven effort of building the envisioned DT.

## 6. CONCLUSION

Even if the goal is often to create a DT that is autonomous and adaptive, the reality is that most DT systems in the present and the foreseeable future will still rely on both humans and computers. Therefore, it becomes necessary to assess how DTs can complement and work with humans. But since DT technology is still in its developmental stage, practitioners and researchers lack this understanding of how DTs integrate with humans. They wonder: *"When humans work with DTs, what types of roles can a DT play, and to what extent can those roles be automated?"* A lack of knowledge about the roles that humans and DTs play in a work system can result in significant costs, misallocation of resources, unrealistic expectations from DTs, and strategic misalignments.

To alleviate this challenge, this paper contributes a two-dimensional conceptual framework, Levels of Digital Twin (LoDT). The framework is an integration of the types of roles a DT can play, broadly categorized under (1) Observer, (2) Analyst, (3) Decision Maker, and (4) Action Executor, and the extent of automation for each of these roles, divided into five different levels ranging from completely manual to fully automated. A particular DT can play any number of roles at varying levels. The framework has been developed and validated by following the DSR research methodology over a time frame of 36 months. The design iterations and the validation were carried out using the feedback from 11 experts over multiple meetings clocking over 50 hours and demonstration sessions with 40+ experts. The framework has also been demonstrated for usefulness by applying it in a real-life case study.

The LoDT framework is intended to be useful for both academia and practice. Educators and researchers could benefit from the new perspective on DTs that emphasizes the change in the agency in a work system from humans to DTs, and therefore, DTs should be viewed as partners fulfilling certain roles and responsibilities and not simply as tools to ease a human's workload. This perspective naturally raises many questions for future research, such as: "How to select the roles that should be delegated to a DT?" "What kind of situations should a DT handle?" and "Who is responsible if a DT makes mistakes?" It also opens many other opportunities for future researchers, including examining social dynamics when DTs work in teams in varied roles, developing models for work delegation between humans and DTs, and researching trust, privacy, and ethics when DTs assume critical roles.

Additionally, the LoDT framework alleviates some of the existing confusion surrounding the DT concept, which is perceived by some researchers to be real-time, continuously updated digital models with predictive and prescriptive capabilities [30,32], while others perceive it as merely a digital representation [33,34]. We believe that none of these definitions are necessarily incorrect. In fact, different researchers have partially proposed different roles that they envision a DT to play; some are more biased towards Observer (therefore emphasizing representation), while others are biased towards Analyst and Decision Maker (emphasizing prediction and prescription). All these definitions unify and make sense when we consider the complete set of roles DTs can perform within the LoDT framework, thus alleviating the confusion surrounding the DT concept. The phenomenon of the importance of "levels" for effective communication and planning has been well documented in other bodies of literature, including the importance of technology readiness levels [104] and levels of autonomous driving [105].

As shown in the case study, practitioners can use the LoDT framework as a tool to (1) plan DT deployments and (2) prepare strategic roadmaps for its future development.

Without a framework, planning DT deployments becomes more of an art than a science as people come up with their own unstructured suggestions about what roles DTs should play and how far they should be automated [4]. Lack of an exhaustive evaluation of all possible options can result in missed opportunities and biases towards a particular technology based on the vision and brilliance of individuals with authority. Through the LoDT framework, structured brainstorming and systematic role allocation between DTs and humans can be conducted to alleviate this issue.

Implementing the most intelligent DT in practice is not a one-shot task. In fact, the creation of a DT is a continuous and evolving process [106]. The LoDT framework, as demonstrated in the case study, can act as a starting point for developing a technology roadmap that helps practitioners to determine the exact level where they currently stand with their DT and, at the same time, establish a long-term vision of where they might want to reach with the DT.

One of the potential areas for future research, as mentioned earlier, is to explore how different role allocations between humans and DTs affect human-DT interfaces and human factors. For the practical deployment of these human-DT systems, it might also be interesting to explore the connected engineering factors such as reliability, robustness, and resilience because they have great implications for humans in terms of safety, health, and well-being [107]. Finally, future researchers can also explore the generalizability of applying this framework in other industries and different technologies.

Ultimately, the pursuit of digitalization and automation requires sustained technological leadership and a clearly articulated vision in the form of a digital strategy. Understanding the different roles that a DT can play, what it takes to achieve them, and the value they provide to the business are critical for developing any strategy. Thus, the LoDT framework should not be viewed as a way to prescribe a particular level of DT to practitioners but rather as a guide for discussing, comparing, communicating, and strategically prioritizing digital opportunities. Awareness of the factors discussed in the framework can help a company stake out a digital strategy that is more likely to succeed.

## DECLARATION OF COMPETING INTERESTS

The authors declare that they have no known competing financial interests or personal relationships that could have appeared to influence the work reported in this paper.

## ACKNOWLEDGEMENTS

We would like to thank all the industry and academic experts who gave us valuable time and feedback. We acknowledge the financial support provided by the Center for Integrated Facility Engineering (CIFE) at Stanford University for completing this project. The authors would also like to express their gratitude to Rui Liu, Stephanie Chin, Melody Spradlin, Cynthia Brosque, Simge Girgin, and Alberto Tono for their feedback on the framework. The authors would also thank the two anonymous reviewers and the editor for their constructive feedback for improving the manuscript. Last but not least, the first author would also like to especially thank Tulika Majumdar, Alissa Cooperman, and Hesam Hamledari for their invaluable feedback and help during framework preparation and manuscript proofreading.